\documentclass[PUB, atlasdraft=false, UKenglish, texmf, orcidlogo]{atlasdoc}
\usepackage{atlaspackage}
\usepackage{atlasbiblatex}
\usepackage{atlascontribute}

\usepackage{atlasphysics}
\graphicspath{{logos/}{figures/}}

\usepackage{ANA-HIGP-2026-18-PUB-defs}

\AtlasTitle{{Lorentz-invariant fiducial decay selection for the ATLAS $H\to ZZ^{*}\to 4\ell$} STXS measurement}

\AtlasAbstract{%
ATLAS measurements of Higgs boson production-mode cross-sections and Simplified Template Cross Sections (STXS), $\sigma\cdot B$, in the $H\to ZZ^{*}\to 4\ell$ decay channel rely on selection factors that fold in a decay-side fiducial acceptance evaluated with Standard Model decay kinematics. In an effective-field-theory interpretation that acceptance depends on the Wilson coefficients themselves. We recast these production-side $\sigma\cdot B$ results into a single Lorentz-invariant fiducial (LIF) decay region defined by the $H\to 4\ell$ decay, absorbing the EFT-dependent fiducial acceptance into the signal parametrisation, while the remaining reconstruction-efficiency corrections are shown to be largely EFT-independent. Using the Higgs boson signal Monte Carlo simulation at $\sqrt{s}=13.6$~TeV, corresponding to an integrated luminosity of $165\,\mathrm{fb}^{-1}$, the LIF acceptance in the Standard Model is found to be $A^{\mathrm{LIF}}_{\mathrm{SM}}=0.8546$, production-independent to $0.07\%$ and stable across STXS bins to $0.19\%$. Less than $1\%$ of the reconstructed signal events originate from outside the LIF region, and the reconstruction efficiency for events in the region changes by at most $6\%$ over the probed ranges of the EFT Wilson coefficients, as in the fiducial phase space of the published analysis. Consequently, the existing inclusive STXS cross-section measurements can be rescaled to the corresponding LIF fiducial phase space, allowing for EFT interpretations without EFT-dependent corrections for the reconstruction efficiency.
}

\AtlasRefCode{PUB-HIGP-2026-18}

\AtlasNote{ATL-PHYS-PUB-2026-029}

\PreprintIdNumber{ATL-PHYS-PUB-2026-029}

\hypersetup{pdftitle={Lorentz-invariant fiducial decay selection for the ATLAS H to ZZ* to 4l STXS measurement},pdfauthor={The ATLAS Collaboration}}

\begin{document}

\maketitle

\section{Introduction}

Simplified Template Cross Sections (STXS) characterise Higgs boson production in bins of particle-level production kinematics. Every bin is defined purely on the production side. Extracting the production cross-section times branching fraction, $\sigma\cdot B$, from the measured signal yield therefore requires knowledge of a selection factor that accounts for both the fiducial acceptance of the decay-side analysis phase space and the efficiency for reconstructing and selecting events within that phase space. These fiducial acceptance and reconstruction efficiency factors are typically evaluated assuming Standard Model (SM) decay kinematics. For interpretations of the measured cross-sections in theories beyond the Standard Model, such as effective field theories (EFTs)~\cite{SMEFT:Warsaw}, this assumption is not necessarily valid. In general, modifications of the decay kinematics can alter both the fiducial acceptance and the reconstruction efficiency, calling for model-dependent corrections that must be evaluated using dedicated EFT simulations.

The published ATLAS $H\to ZZ^{*}\to4\ell$ measurement at $\sqrt{s}=13.6$~\TeV reports $\sigma_{|\yH|<2.5}\cdot B(H\to ZZ^{*})$ for the Stage-0 production modes and for the subset of Reduced Stage-1.2 bins that the data constrain~\cite{HIGP-2025-09}. This note recasts those results into a \emph{Lorentz-invariant fiducial (LIF) decay region}: a particle-level selection of the four-lepton final state built from invariant observables only, with no requirement on individual lepton momenta or directions. Initial studies of such a simplified fiducial decay volume were reported in Ref.~\cite{ATLAS:simplefid}. Each published cross-section is multiplied by the SM value of a fiducial decay acceptance, $\AsfSM$, evaluated at particle level from the Higgs boson signal Monte Carlo (MC) simulation samples used in Ref.~\cite{HIGP-2025-09}. \Cref{sec:strategy} defines the fiducial selection criteria for the LIF phase space, the fiducial acceptance $\Asf$, and two reconstruction-related ratios used to assess the model independence of the approach. \Cref{sec:acceptance} presents the values of these quantities for the different SM Higgs boson production processes. Their dependence on the relevant EFT Wilson coefficients is investigated in \cref{sec:modelindep}. The resulting rescaled cross-sections are presented in \cref{sec:rescale}. The $\AsfSM$ factors derived from Run-3 simulation are also applied to the Run-2 STXS measurements~\cite{HIGG-2018-28}, since no significant variation in the decay-side acceptance is expected between the two data-taking periods.

\section{Rescaling strategy}
\label{sec:strategy}

The LIF selection uses Lorentz-invariant four-lepton observables only, so the boost of the four-lepton system does not affect the associated acceptance. For a spin-0 Higgs boson, production and decay are expected to factorise. Together these two statements imply that the acceptance of the LIF region is a single number: the same for every production mode, and independent of the $N_{\mathrm{jets}}$, \ptH\ and $m_{jj}$ requirements that define the STXS bins.  \Cref{sec:acceptance} tests that expectation rather
than assuming it.

\Cref{tab:fiddef} summarises the LIF selection criteria, together with those of the nominal fiducial phase space of the published analysis (Table~2 of Ref.~\cite{HIGP-2025-09}). They differ from the nominal fiducial phase space mainly in that the charged-lepton requirements $\pt > 20,\ 15,\ 10\,\GeV$ and $|\eta|<2.7$ are omitted. All comparisons below are made against that fiducial phase space definition.

Leptons are defined as electrons and muons that do not originate from hadron or $\tau$-lepton decays. They are dressed
by adding the four-momenta of photons whose opening angle to the lepton in the Higgs boson rest frame is below $0.1$\,rad, excluding
photons from hadron decays, so that energy lost to final-state radiation is partially recovered. Photons radiated at larger angles are not included, as in the nominal fiducial phase space. Only leptons from the Higgs boson decay are used; leptons from the decays of vector bosons or top quarks produced in association with the Higgs boson are ignored, so no matrix-element-based choice of the lepton quadruplet is needed.

\begin{table}[!ht]
\centering
\small
\caption{Definition of the LIF selection, evaluated at particle level, and of the nominal fiducial phase space of Ref.~\cite{HIGP-2025-09}. As in the published analysis, the leading pair is the same-flavour, opposite-charge (SFOC) lepton pair with invariant mass $m_{12}$ closest to the $Z$-boson mass, and the subleading pair is the remaining SFOC lepton pair with invariant mass $m_{34}$ closest to the $Z$-boson mass. The LIF selection uses invariant
four-lepton observables only and applies no \pt\ or $\eta$ requirement on charged leptons. A dash denotes no requirement.}
\label{tab:fiddef}
\begin{tabular}{>{\raggedright\arraybackslash}p{0.19\textwidth}>{\raggedright\arraybackslash}p{0.34\textwidth}>{\raggedright\arraybackslash}p{0.36\textwidth}}
\toprule
Requirement            & LIF region & Nominal fiducial phase space \\
\midrule
Lepton kinematics      & --- & $\pt > 5\,\GeV$ and $|\eta| < 2.7$; $\pt > 20,\ 15,\ 10\,\GeV$ for the three leading leptons \\
Mass requirements      & $50 < m_{12} < 106\,\GeV$ and $12 < m_{34} < 115\,\GeV$ & $50 < m_{12} < 106\,\GeV$ and $12 < m_{34} < 115\,\GeV$ \\
Lepton separation      & opening angle between two leptons in the Higgs boson rest frame $> 0.1$\,rad & $\Delta R(\ell_i,\ell_j) > 0.1$ \\
Lepton--jet separation & --- & $\Delta R(\ell_i,\mathrm{jet}) > 0.1$ \\
Low-mass dilepton veto & $m(\ell_i,\ell_j) > 5\,\GeV$ for all SFOC pairs & $m(\ell_i,\ell_j) > 5\,\GeV$ for all SFOC pairs \\
Additional leptons     & only leptons from the Higgs boson decay are used & quadruplet with the largest matrix-element value if an extra lepton with $\pt > 12\,\GeV$ is present \\
\bottomrule
\end{tabular}
\end{table}

For each STXS production bin $i$ the particle-level fiducial acceptance is
\begin{equation}
\Asf_i \;=\;
\frac{\sum_{j\in\text{bin }i}\, w_j\,\big[\text{decay in the LIF region}\big]}{\sum_{j\in\text{bin }i}\, w_j\,\big[\text{all }H\to 4\ell\big]}\,,
\label{eq:asf}
\end{equation}
where the sums run over generated signal events $j$ of weight $w_j$ at particle level. These STXS bins comprise the Stage-0 modes (ggF, VBF, VH) and, at finer
granularity, the eleven populated Reduced Stage-1.2 bins. Each STXS production bin defines a given production
phase space, whereas $\Asf$ measures only the decay. In an EFT, the decay kinematics and hence the fiducial acceptance depend on the
Wilson coefficients $c$, $\Asf_i=\Asf_i(c)$. Its SM value, $\AsfSMbin{i}\equiv\Asf_i(c=0)$, is evaluated with the SM
simulation and used to rescale the published cross-section:
\begin{equation}
\Ssf_i \;=\; S_i \times \frac{B(H\to ZZ^{*}\to 4\ell)}{B(H\to ZZ^{*})} \times \AsfSMbin{i}\,,
\label{eq:rescale}
\end{equation}
with $S_i \equiv \sigma_{|\yH|<2.5}\cdot B(H\to ZZ^{*})$ the published value and the branching
ratios taken from Ref.~\cite{LHCHWG:YR4}. The ratio converts a cross-section quoted for
$H\to ZZ^{*}$ into one for the four-lepton final state before the fiducial acceptance is applied. Here $\Asf$ is the
particle-level acceptance $a_j$ of Equation~(3) of Ref.~\cite{HIGP-2025-09}, redefined on invariant
decay observables.

The fiducial acceptance $\Asfc$ itself, as well as the overall acceptance of reconstructed events, changes with the EFT parameters. The dependence of $\Asfc$ is part of the EFT prediction for the LIF region, which is computed at particle level, and is quantified in \cref{sec:modelindep}. In order to employ the LIF-rescaled cross-sections for EFT interpretations, it is necessary to demonstrate that the number of events in the LIF region depends on the EFT parameters in essentially the same way as the number of reconstructed events, i.e. that the efficiency for reconstructing events in the LIF region and the fraction of reconstructed events originating outside it do not exhibit a significant dependence on the EFT parameters. For that purpose, two reconstruction-related ratios are defined:
\begin{equation}
\rfid \;=\; \frac{N_{\mathrm{fid}}}{N_{\mathrm{fid}\,\&\,\mathrm{reco}}}\,,
\qquad
\rreco \;=\; \frac{N_{\mathrm{reco}}}{N_{\mathrm{fid}\,\&\,\mathrm{reco}}}\,.
\label{eq:ratios}
\end{equation}
Here $N_{\mathrm{fid}}$ counts events whose particle-level decay passes a given fiducial selection,
whether or not they are reconstructed. $N_{\mathrm{reco}}$ counts events passing the
reconstruction-level selection, whether or not their particle-level decay lies inside the fiducial region. And
$N_{\mathrm{fid}\,\&\,\mathrm{reco}}$ counts those satisfying both. The two ratios share a
denominator and are both $\ge1$. The first, $\rfid$, is the inverse of the reconstruction efficiency within the fiducial region. The second, $\rreco-1$, is the fraction of the reconstructed
signal-region yield produced outside it, and corresponds to the fiducial leakage factor
$f^{\mathrm{nonfid}}$ of Ref.~\cite{HIGP-2025-09} evaluated for the LIF region.

With $N_{\mathrm{incl}}$ the number of all $H\to4\ell$ decays, the overall acceptance of reconstructed events, $A^{\mathrm{reco}}=N_{\mathrm{reco}}/N_{\mathrm{incl}}$, factorises into the fiducial acceptance and the reconstruction efficiency, $A^{\mathrm{reco}}=\Asf\cdot\rreco/\rfid$, with $\rfid$ and $\rreco$ evaluated for the LIF region. The published cross-sections are extracted with selection factors evaluated in the SM. For a signal modified by Wilson coefficients $c$, a published value $S$ therefore corresponds to $S\cdot B(H\to ZZ^{*}\to 4\ell)/B(H\to ZZ^{*})=\sigma(c)\,B_{4\ell}(c)\,A^{\mathrm{reco}}(c)/A^{\mathrm{reco}}_{\mathrm{SM}}$, where $\sigma(c)\,B_{4\ell}(c)$ is the predicted cross-section times branching fraction to four leptons. The rescaled cross-section of \cref{eq:rescale} then becomes
\begin{equation}
\Ssf \;=\; \sigma(c)\,B_{4\ell}(c)\,\Asfc\times\frac{\rreco(c)}{\rreco(\mathrm{SM})}\times\frac{\rfid(\mathrm{SM})}{\rfid(c)}\,.
\label{eq:eftrelation}
\end{equation}
The EFT-dependent fiducial acceptance thus enters through the signal prediction for the LIF region, $\sigma(c)\,B_{4\ell}(c)\,\Asfc$, which is computed at particle level. The last two factors describe the reconstruction efficiency, and \cref{sec:modelindep} shows that they remain close to unity.

\section{Lorentz-invariant fiducial acceptance and out-of-fiducial fractions}
\label{sec:acceptance}

To check the purity of the LIF region, its requirements are applied to the decay products of the Higgs boson in all
decay modes. \Cref{tab:decaymodes} follows an inclusive sample of $10^{6}$ Higgs boson decays,
classified by the immediate children of the Higgs boson, through the requirements of \cref{tab:fiddef} applied in
sequence, followed by the four-lepton mass range of the Higgs boson signal region, $115 < \mfl < 130\,\GeV$. After the $m_{34}$ requirement only $H\to ZZ^{*}\to4\ell$ and two $H\to b\bar b$ events
survive, and after the mass range only $H\to ZZ^{*}\to4\ell$ events remain. The LIF region is therefore pure in $H\to ZZ^{*}\to4\ell$ events at particle level, and no correction for
other decay modes enters the rescaling.

\begin{table}[htbp]
\centering
\footnotesize
\caption{Number of events passing each LIF requirement in an inclusive sample of
$10^{6}$ Higgs boson decays, applied sequentially. The decay mode is that of the immediate children
of the Higgs boson. The last column applies the four-lepton mass range of the Higgs boson signal region,
$115 < \mfl < 130\,\GeV$. Empty entries denote no surviving events.}
\label{tab:decaymodes}
\begin{tabular}{lrrrrrrr}
\toprule
Decay mode & Inclusive & $N_\ell = 4$ & $m_{ij}$ veto & $m_{12}$ & $m_{34}$ & angle & $\mfl$ \\
\midrule
$ZZ^{*}\to4\ell$      & 119      & 119     & 117 & 115 & 110 & 110 & 108 \\
$ZZ^{*}\to\nu\nu\ell\ell$ & 1\,833   & 17      & 9   & 3   & --- & --- & --- \\
$ZZ^{*}\to jj\ell\ell$ & 24\,267  & 837     & 79  & 30  & --- & --- & --- \\
$WW$                  & 215\,973 & 2\,246  & 6   & --- & --- & --- & --- \\
$b\bar b$             & 577\,010 & 19\,389 & 326 & 4   & 2   & 1   & --- \\
$c\bar c$             & 28\,919  & 368     & 1   & --- & --- & --- & --- \\
$gg$                  & 84\,874  & 1\,736  & 1   & --- & --- & --- & --- \\
$\tau\tau$            & 62\,731  & 32      & --- & --- & --- & --- & --- \\
$\gamma\gamma$        & 2\,295   & 1       & --- & --- & --- & --- & --- \\
$Z\gamma$             & 1\,573   & 19      & --- & --- & --- & --- & --- \\
$\mu\mu$              & 211      & ---     & --- & --- & --- & --- & --- \\
$s\bar s$             & 193      & ---     & --- & --- & --- & --- & --- \\
$u\bar u$, $d\bar d$  & 2        & ---     & --- & --- & --- & --- & --- \\
\bottomrule
\end{tabular}
\end{table}

\begin{figure}[htbp]
\centering
\includegraphics[width=0.70\textwidth]{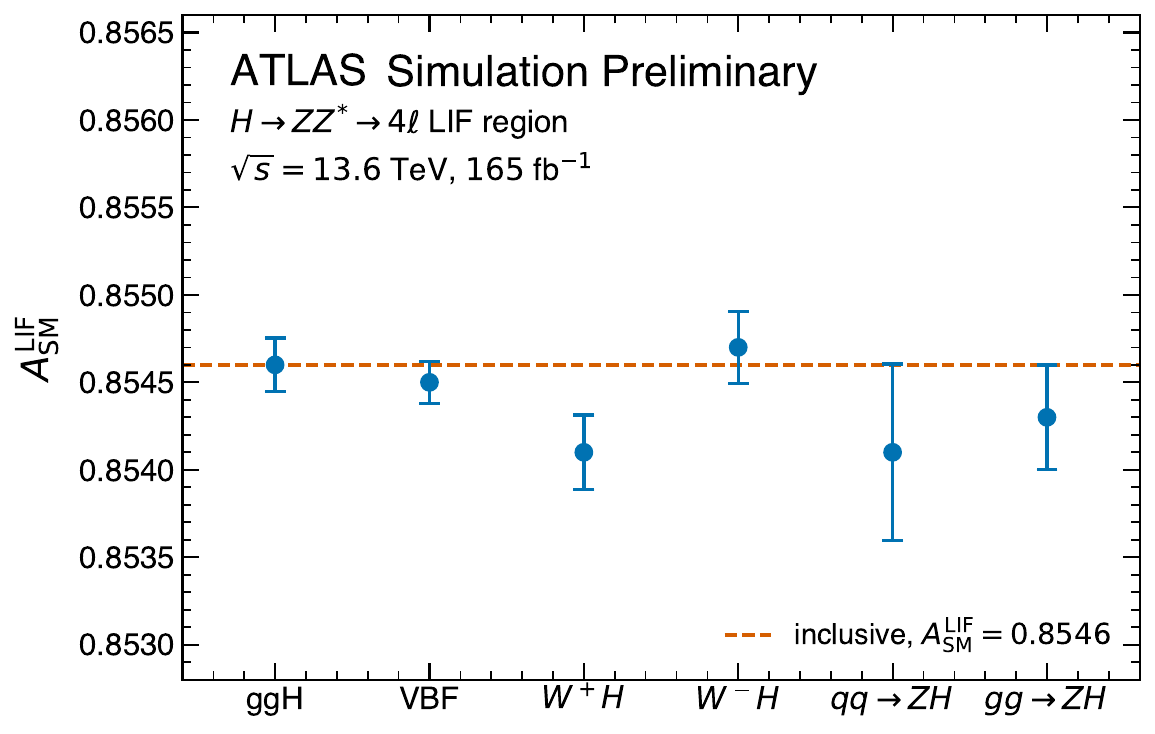}
\caption{The Lorentz-invariant fiducial acceptance in the SM, $\AsfSM$, for each of the six Higgs boson signal
samples. The error bars represent the weighted binomial uncertainty from the number of simulated events. The dashed
line indicates the inclusive value, obtained from the six samples combined according to their cross-sections.}
\label{fig:lif_samples}
\end{figure}

The SM acceptance $\AsfSM$ of \cref{eq:asf} is evaluated on the Higgs boson signal MC simulation for the ggF, VBF and $VH$ production modes: the ggH, VBF, $W^+H$, $W^-H$, $qq\to ZH$ and $gg\to ZH$ samples of Ref.~\cite{HIGP-2025-09}, generated with \textsc{Powheg Box}~v2~\cite{Nason:2004rx,Frixione:2007vw,Alioli:2010xd} and interfaced to \textsc{Pythia}~8~\cite{Bierlich:2022pfr}. The evaluation uses $8.3$, $8.9$, $3.0$, $3.2$, $3.1$ and $1.4$ million simulated events from these samples, respectively ($28.0$ million in total). Inclusively, a value of $\AsfSM=0.8546$ is obtained. The values obtained from the individual samples are shown in Figure~\ref{fig:lif_samples}.

\begin{figure}[htbp]
\centering
\includegraphics[width=0.80\textwidth]{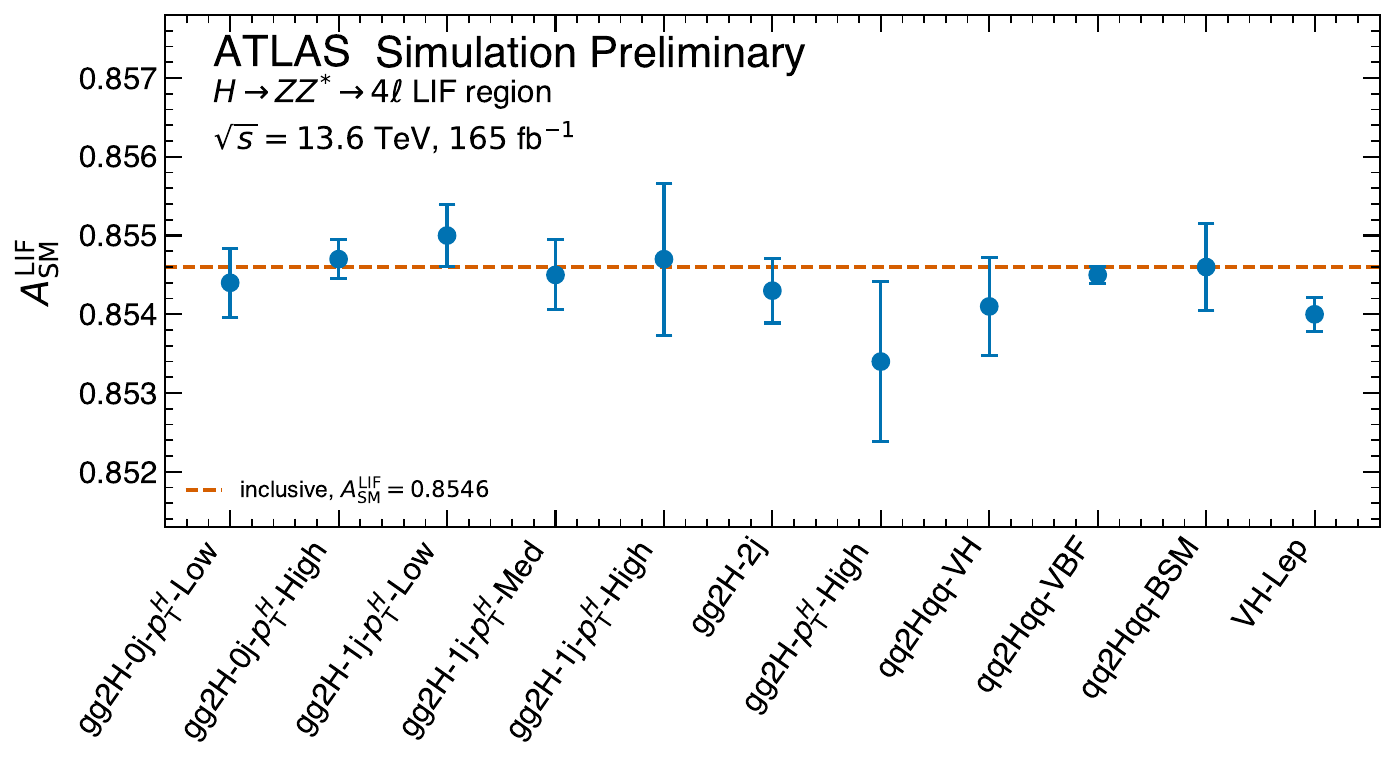}
\caption{The Lorentz-invariant fiducial acceptance in the SM, $\AsfSM$, in the eleven populated Reduced Stage-1.2
bins. The error bars represent the weighted binomial uncertainty from the number of simulated events. The dashed
line indicates the inclusive value, obtained from the six samples combined according to their cross-sections.}
\label{fig:lif_reduced}
\end{figure}

The six samples span $\AsfSM$ values from $0.8541$ to $0.8547$, with a relative spread of $0.07\%$. Grouped into Stage-0 categories the values are $0.8546$ (ggF), $0.8545$ (VBF) and $0.8543$ (VH), with a $0.04\%$ spread.
Across the eleven populated Reduced Stage-1.2 bins these span the range of values from $0.8534$ to $0.8550$, with a $0.19\%$ spread, as shown in Figure~\ref{fig:lif_reduced}. The weighted binomial uncertainties of these acceptances,
$\sigma^2(\Asf)=\big[(1-\Asf)^{2}\sum_{\mathrm{pass}}w_i^2+{\Asf}^{2}\sum_{\mathrm{fail}}w_i^2\big]
\big/\big(\sum_i w_i\big)^{2}$, range from $0.015\%$ to $0.023\%$ relative at Stage-0 and from $0.013\%$ to $0.12\%$
across the Reduced Stage-1.2 bins.

\begin{figure}[htbp]
\centering
\includegraphics[width=0.55\textwidth]{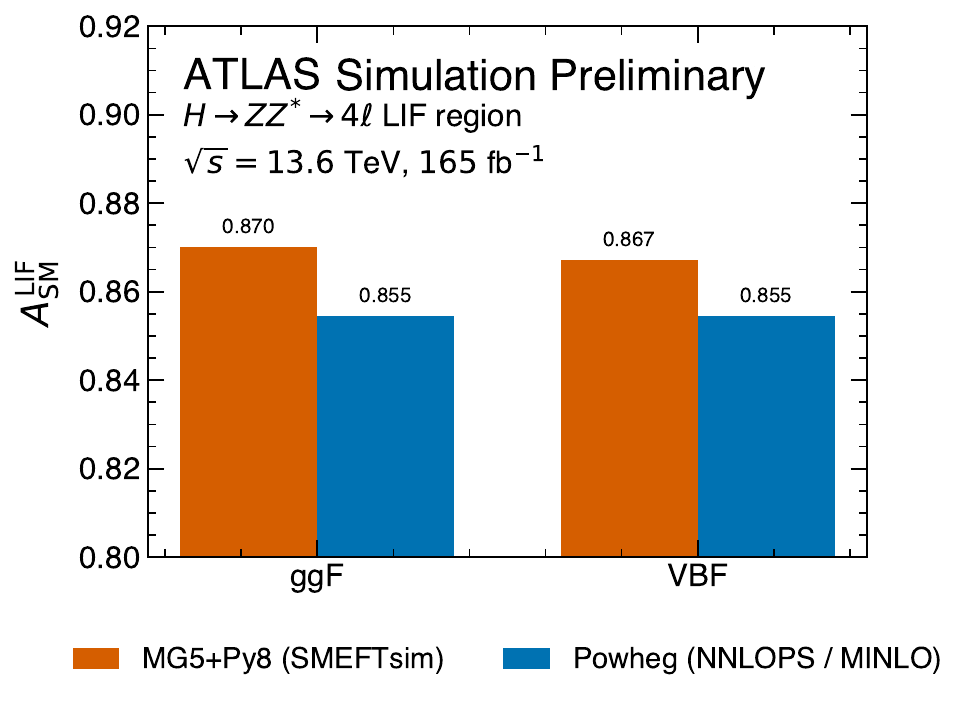}
\caption{Comparison of the Lorentz-invariant fiducial acceptance in the SM, $\AsfSM$, obtained from the
MadGraph5\_aMC@NLO$+$\allowbreak\textsc{Pythia}8 SMEFTsim samples with that obtained from the \textsc{Powheg}-based Higgs boson signal
simulation, for ggF and VBF production.}
\label{fig:lif_generator}
\end{figure}

The $\AsfSM$ values for $VH$ production with leptonically decaying vector bosons are $0.8542$, $0.8537$ and $0.8537$ for $qq\to WH$, $qq\to ZH$ and $gg\to ZH$ production, respectively. Replacing the
MadGraph5\_aMC@NLO$+$\textsc{Pythia}8 SMEFTsim samples~\cite{Alwall:2014hca,Bierlich:2022pfr,Brivio:2020onw} described in \cref{sec:modelindep} by the \textsc{Powheg}-based Higgs boson signal simulation
shifts $\AsfSM$ by $-1.79\%$ for ggF and $-1.47\%$ for VBF; the same generator replacement changes the acceptance value in the nominal fiducial selection by about $7\%$. The acceptances from the two simulations are compared in Figure~\ref{fig:lif_generator}. This difference of less than $2\%$ is taken as an estimate of the acceptance-modelling uncertainty when extrapolating to the inclusive decay phase space and is therefore not relevant for the measurements in the LIF phase space presented here.

\begin{figure}[htbp]
\centering
\includegraphics[width=0.80\textwidth]{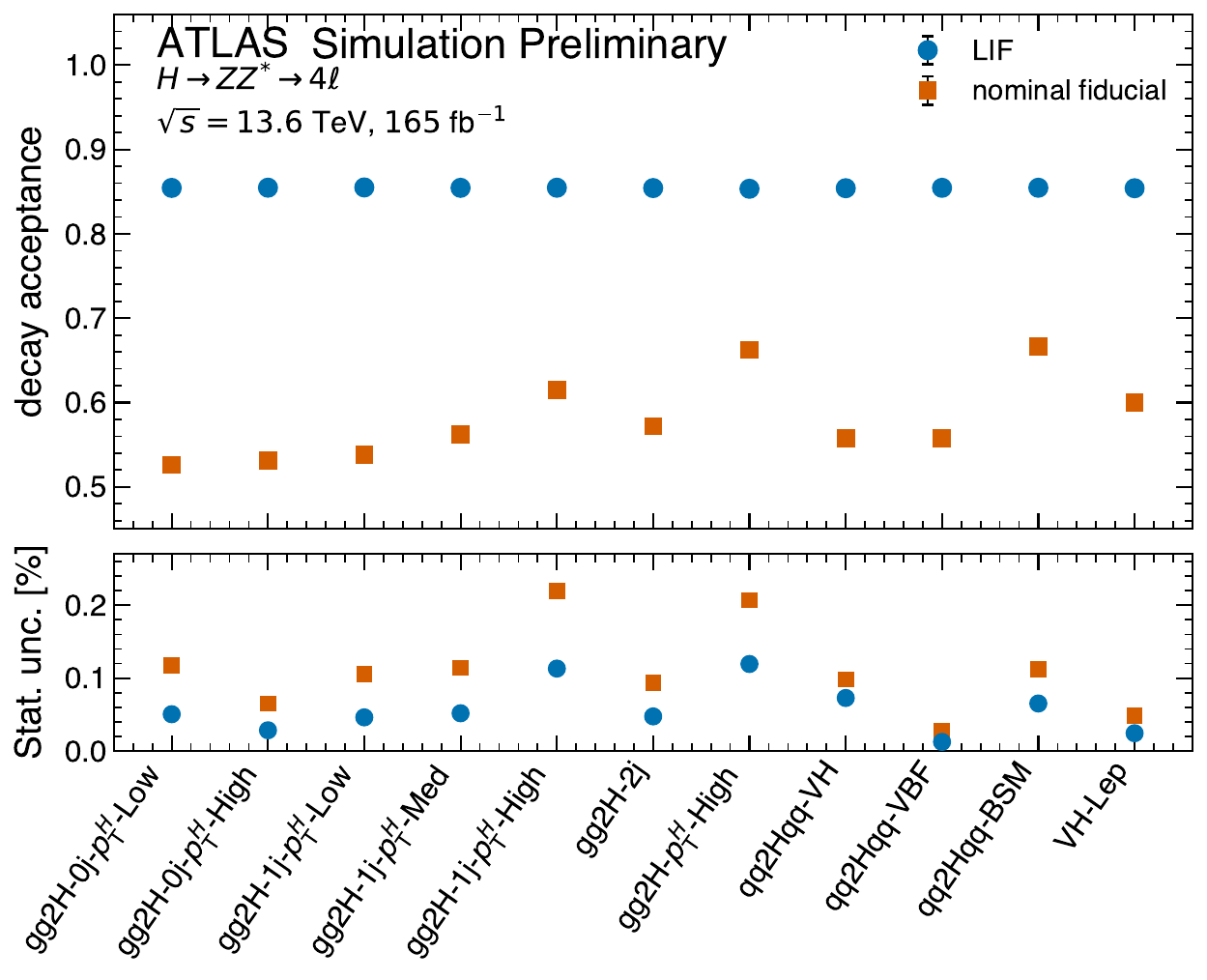}
\caption{Comparison of the decay acceptance of the Lorentz-invariant fiducial region (circles) with that of the
nominal fiducial phase space of Ref.~\cite{HIGP-2025-09} (squares), in the eleven populated Reduced Stage-1.2 bins.
The error bars represent the weighted binomial uncertainty from the number of simulated events; the uncertainty is smaller
than the marker size and, thus, not visible. The lower panel shows the relative size of this uncertainty for each point.}
\label{fig:lif_vs_nominal}
\end{figure}

This behaviour is consistent with the expectations outlined in \cref{sec:strategy}: no dependence on the production mode and production kinematics is observed within statistical uncertainties. The contrast with the nominal fiducial phase space of Ref.~\cite{HIGP-2025-09} is large: the nominal fiducial acceptance ranges from $0.526$ to $0.666$, with a maximum-to-minimum ratio of about $1.27$. This is primarily driven by the
lepton \pt\ selection requirements, whose efficiency is affected by the boost of the four-lepton system. The acceptances from the nominal fiducial and the LIF selections are compared in Figure~\ref{fig:lif_vs_nominal}.

A particle-level event counts as reconstructed when the same event also passes the reconstruction-level selection. With the reconstruction restricted to the Higgs boson signal region, $115 < \mfl < 130\,\GeV$,
$\rreco-1$ ranges from $0.4\%$ to $0.5\%$ for the process-integrated ggF, VBF and VH samples, and stays below
$0.6\%$ for each individual VH channel. For individual reconstructed event categories, the values are spread from $0.3\%$ to $0.5\%$ in
categories with high signal yield, and from $1.1\%$ to $1.4\%$ in the sparsely populated
leptonic $VH$ categories. In each reconstructed event category, the value of $\rreco-1$ obtained from the events of the dominant STXS production bin alone agrees to within $0.0015$ with that obtained from all
production bins contributing to the category, so the dependence on the production composition of the category is weak.

\section{Assessment of the EFT model dependence}
\label{sec:modelindep}

Using $\AsfSM$ in an EFT interpretation requires two validation tests. First, the reconstructed signal outside the LIF region ($\rreco-1$) should be small, so that events outside the region contribute little to the reconstructed yield. Second, the ratios $\rfid$ and $\rreco$ of \cref{eq:ratios} should not change significantly when a Wilson coefficient is varied. The size of the out-of-fiducial contribution is evaluated with the SM Higgs boson signal MC simulation in \cref{sec:acceptance}. Those simulations carry no per-event EFT weights. Instead, the EFT dependence is evaluated with dedicated EFT-reweighted ggF and VBF signal samples, where the Wilson coefficients are defined in the Warsaw basis~\cite{SMEFT:Warsaw}. The samples were generated at leading order (LO) in QCD with MadGraph5\_aMC@NLO~3.5.7~\cite{Alwall:2014hca}, using the SMEFTsim~3 model~\cite{Brivio:2020onw} (top flavour scheme, $\{m_W, m_Z, G_F\}$ input scheme) and the NNPDF3.0nlo PDF set~\cite{Ball:2014uwa}, and were interfaced to \textsc{Pythia}~8.308~\cite{Bierlich:2022pfr} for the parton shower and hadronisation, with \textsc{EvtGen}~2.2.1~\cite{Lange:2001uf} for the decays of bottom and charm hadrons. The conclusions below therefore hold for the sampled production modes,
operators and one-coefficient scan ranges.

Six EFT operators associated with the $HVV$ interaction vertex are probed, one at a time: the CP-even $c_{HW}$, $c_{HB}$ and $c_{HWB}$ and the CP-odd $c_{H\widetilde{W}}$, $c_{H\widetilde{B}}$ and $c_{H\widetilde{W}B}$, each scanned over $-2\le c\le 2$~\TeV$^{-2}$ with the others fixed to zero. As reference intervals, the marginalised $95\%$~CL intervals of the global EFT fit of Ref.~\cite{Ellis:2020unq} are used for the CP-even coefficients, and $-1\le c\le 1$~\TeV$^{-2}$ for the CP-odd coefficients, which are not included in that fit. The largest relative change of $\rreco$ obtained from these scans is $0.6\%$. The
reconstructed-to-fiducial correspondence is thus approximately independent of the Wilson coefficient value within the
scanned range. \Cref{tab:eft_stability} compares the obtained $\rfid$ values for the three different phase-space regions (LIF, nominal fiducial and inclusive). The values for the LIF region and the nominal fiducial phase space agree with each other to better than $1\%$ at every representative Wilson coefficient value.
Over the reference intervals of the Wilson coefficients, the largest deviation from the SM ranges from $0.4\%$ to $6.1\%$ ($0.4\%$ to $5.5\%$) for the LIF region (nominal fiducial phase space), so the EFT dependence of the LIF region follows that of the reconstructed events to within a few percent. For the inclusive phase space without any fiducial requirement, the two differ much more, with deviations ranging
from $4.2\%$ for $c_{HB}$ up to $93\%$ for $c_{H\widetilde{B}}$. This inclusive selection is the one implicitly associated with the published STXS cross-section measurements. The LIF region restores a level of EFT independence comparable to that of the nominal fiducial phase space, while removing the dependence on the production bin.
Requiring a fiducial region of either kind is therefore what stabilises the mapping. Removing the lepton momentum and direction requirements, as done in the LIF region, does not significantly degrade the stability within the scanned parameter range.

The fiducial acceptance itself depends on the Wilson coefficients, as shown by the $\Asfc/\AsfSM$ column of \cref{tab:eft_stability}, which is evaluated on the same samples. Over the reference intervals, $\Asfc$ departs from $\AsfSM$ by between $3.7\%$ for $c_{HB}$ and $46\%$ for $c_{H\widetilde{B}}$, because the modified decay kinematics move events into or out of the LIF region. The ratio of the inclusive $\rfid$ in the EFT to its SM value is related to the LIF quantities by
\begin{equation}
\frac{\rfid^{\mathrm{incl}}(c)}{\rfid^{\mathrm{incl}}(\mathrm{SM})} \;=\; \frac{\rfid^{\mathrm{LIF}}(c)}{\rfid^{\mathrm{LIF}}(\mathrm{SM})}\times\left(\frac{\Asfc}{\AsfSM}\right)^{-1}\times\left(\frac{\rreco^{\mathrm{LIF}}(c)}{\rreco^{\mathrm{LIF}}(\mathrm{SM})}\right)^{-1}.
\end{equation}
The large EFT-induced departures of the inclusive selection are therefore driven primarily by the fiducial-acceptance ratio $\Asfc/\AsfSM$, which is part of the EFT signal prediction for the LIF region, $\sigma(c)\,B_{4\ell}(c)\,\Asfc$, while the residual EFT dependence of $\rfid^{\mathrm{LIF}}$ and $\rreco^{\mathrm{LIF}}$ is small (\cref{tab:eft_stability}).

The recasting of the published STXS cross-section measurements is therefore performed by scaling them with $\AsfSM$,
and the reconstruction-efficiency factors of \cref{eq:eftrelation} do not need to be modelled in an EFT-dependent way. Two limitations bound this approximation. First, $\rreco-1$ describes only the LIF region defined here. It says nothing about the published nominal-fiducial leakage, reported
separately at the $1$--$2\%$ level and rising to about $6\%$ in the last $m_{34}$
bin~\cite{HIGP-2025-09}. Second, the departure of $\rfid(c)/\rfid(\mathrm{SM})$ depends on both the coefficient
and the production mode, and propagating it into a combined fit requires the observed covariance of the published measurement, which is not part of the public result; it is left to a subsequent interpretation that has access to the likelihood.

\begin{table}[htbp]
\centering
\footnotesize
\setlength{\tabcolsep}{4pt}
\caption{EFT stability of the fiducial ratio at representative points. Each coefficient is scanned over \mbox{$-2\le c\le 2$~\TeV$^{-2}$} with all others fixed to zero, and all quantities are evaluated at the displayed value. The
third column gives the change of the overall acceptance, $A^{\mathrm{reco}}(c)/A^{\mathrm{reco}}_{\mathrm{SM}}$, with $A^{\mathrm{reco}}=N_{\mathrm{reco}}/N_{\mathrm{incl}}$ and $N_{\mathrm{incl}}$ the number of all $H\to4\ell$ decays, and the
fourth column that of the fiducial acceptance, $\Asfc/\AsfSM$. Both acceptances are normalised to $N_{\mathrm{incl}}$, so the change of the total $H\to4\ell$ rate cancels and the two columns can be compared directly. The next three columns give $\rfid$ of \cref{eq:ratios} for three
selections: the LIF region, the nominal fiducial phase space, and no fiducial requirement
(Inclusive), the last corresponding to the published selection factors. Since every reconstructed event lies in the inclusive phase space, the Inclusive column is the inverse of the third column. Entries are for the ggF sample;
VBF values differ by less than $4\%$ for $A^{\mathrm{reco}}(c)/A^{\mathrm{reco}}_{\mathrm{SM}}$, by less than $6\%$ for
$\Asfc/\AsfSM$ and by less than $2\%$ for $\rfid$.
The last column is the maximum $|\rreco^{\mathrm{EFT}}/\rreco^{\mathrm{SM}}-1|$ over the full
scanned range.}
\label{tab:eft_stability}
\begin{tabular}{lccccccc}
\toprule
& & & & \multicolumn{3}{c}{$\rfid^{\mathrm{EFT}}/\rfid^{\mathrm{SM}}$} & \\
\cmidrule(lr){5-7}
Coefficient & Value (\TeV$^{-2}$) & $A^{\mathrm{reco}}(c)/A^{\mathrm{reco}}_{\mathrm{SM}}$ & $\Asfc/\AsfSM$ & LIF & Nominal & Inclusive & $\max|\rreco^{\mathrm{EFT}}/\rreco^{\mathrm{SM}}-1|$ \\
\midrule
$c_{HW}$              & $+0.6$ & $0.903$ & $0.909$ & $1.007$ & $1.006$ & $1.107$ & $0.35\%$ \\
$c_{HWB}$             & $+0.7$ & $0.867$ & $0.870$ & $1.004$ & $1.003$ & $1.153$ & $0.26\%$ \\
$c_{HB}$              & $+0.2$ & $0.960$ & $0.963$ & $1.003$ & $1.003$ & $1.041$ & $0.50\%$ \\
$c_{H\widetilde{W}}$  & $+1.0$ & $0.781$ & $0.788$ & $1.011$ & $1.008$ & $1.281$ & $0.35\%$ \\
$c_{H\widetilde{B}}$  & $+1.0$ & $0.519$ & $0.541$ & $1.045$ & $1.038$ & $1.927$ & $0.60\%$ \\
$c_{H\widetilde{W}B}$ & $+1.0$ & $0.752$ & $0.764$ & $1.017$ & $1.014$ & $1.330$ & $0.26\%$ \\
\bottomrule
\end{tabular}
\end{table}

\section{Rescaled Lorentz-invariant fiducial cross-sections}
\label{sec:rescale}

\begin{figure}[htbp]
\centering
\includegraphics[width=0.70\textwidth]{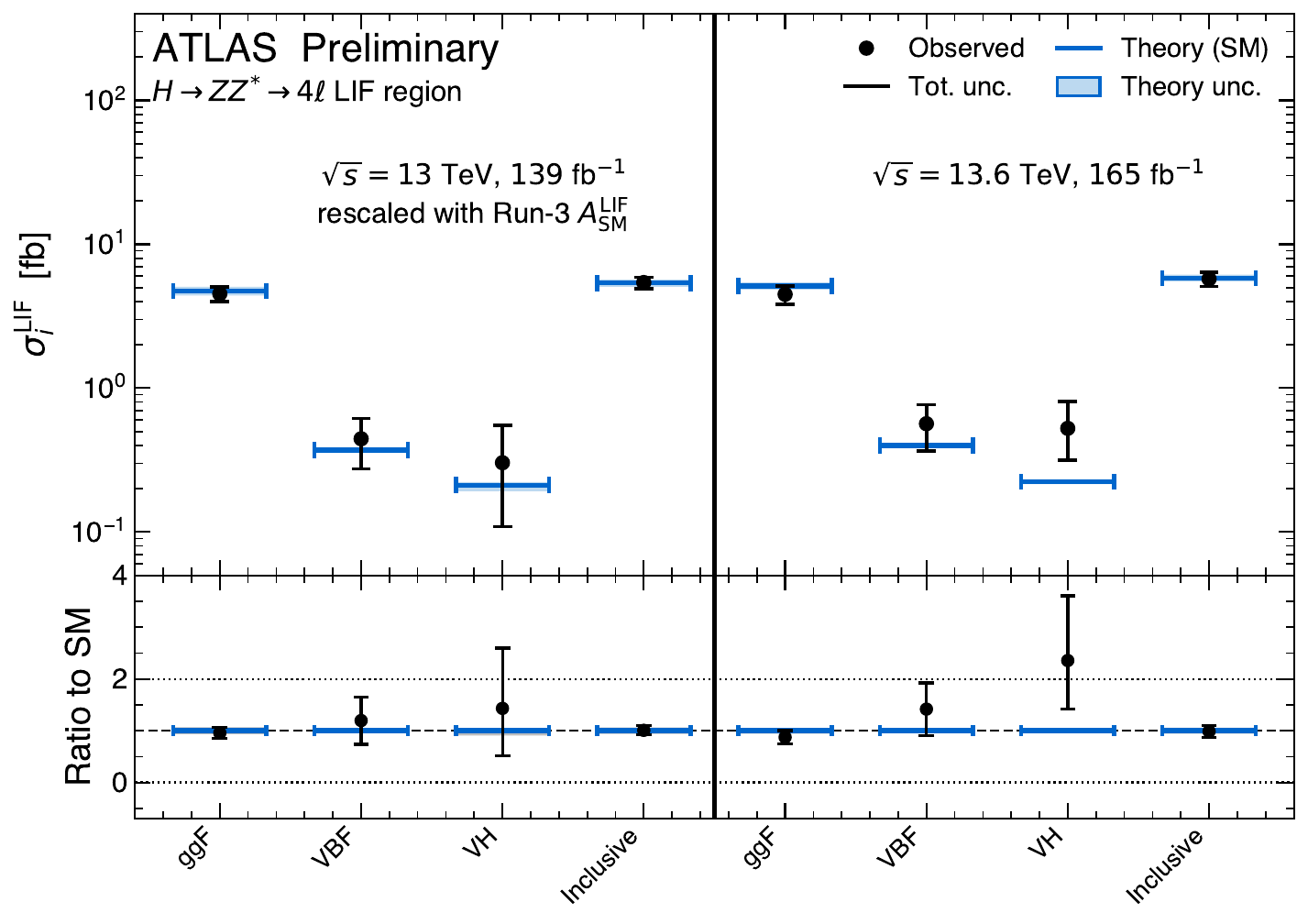}
\caption{Rescaled Lorentz-invariant fiducial cross-sections $\Ssf_i$ for the production-mode
categories, for the Run-2 measurement of Ref.~\cite{HIGG-2018-28} (left of the vertical line) and the Run-3
measurement of Ref.~\cite{HIGP-2025-09} (right). The markers show the observed values and the error bars the total
uncertainty in the published input cross-sections. The horizontal bars show the Standard Model predictions of the
same references, rescaled by the same factor, and the shaded bands the uncertainty in those predictions; where this uncertainty is small, the band is hidden by the bar. The Run-2 cross-sections are rescaled with the $\AsfSM$ values obtained from the Run-3 simulation. The lower
panel shows the ratio of the observed values to the Standard Model predictions, with the uncertainty in the
prediction shown separately as the grey band around unity.}
\label{fig:slif_stage0}
\end{figure}

\begin{figure}[htbp]
\centering
\includegraphics[width=\textwidth]{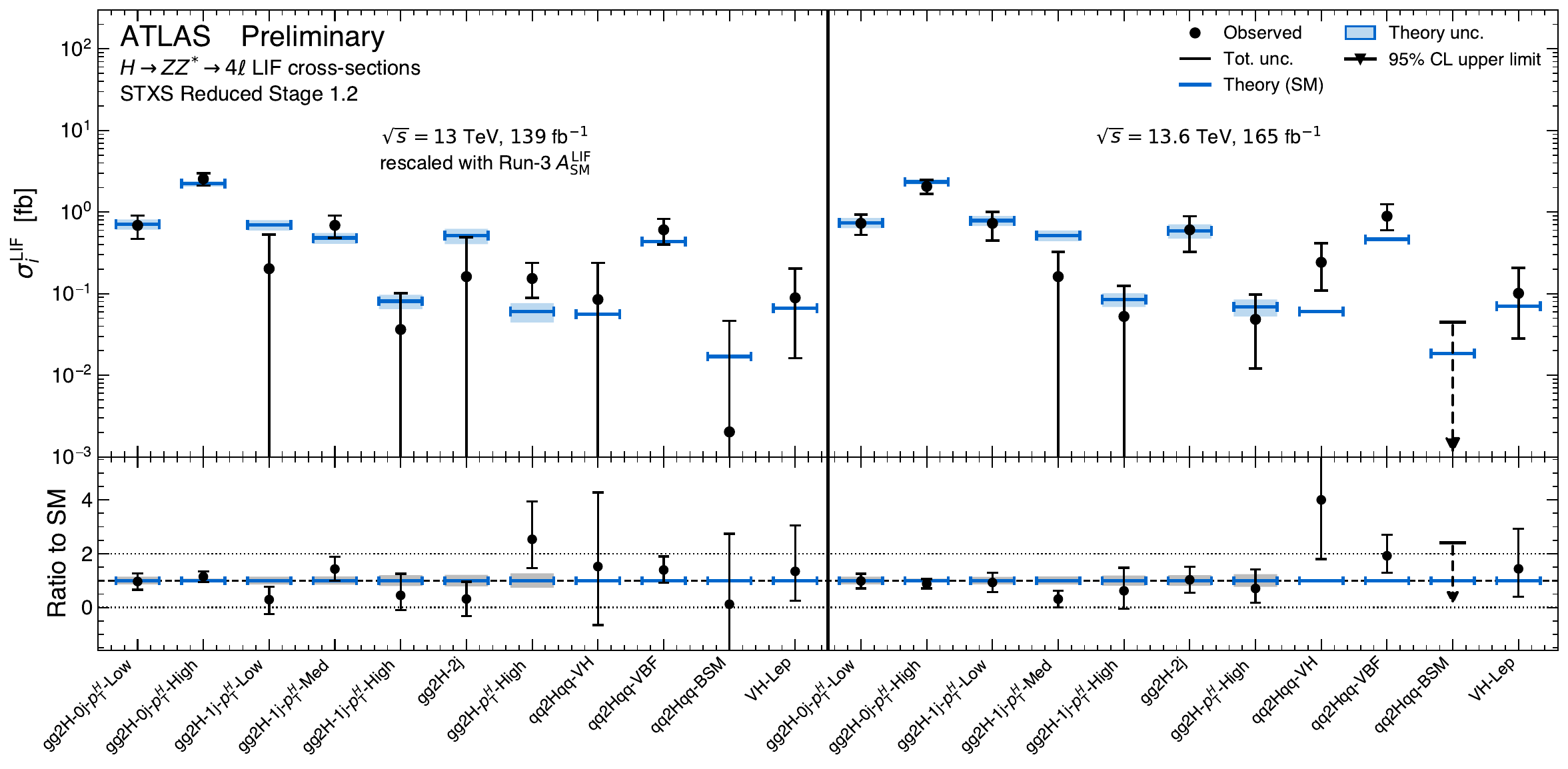}
\caption{Rescaled Lorentz-invariant fiducial cross-sections $\Ssf_i$ for the STXS Reduced Stage-1.2
bins of the Run-2 measurement of Ref.~\cite{HIGG-2018-28} (left of the vertical line) and of the
Run-3 measurement of Ref.~\cite{HIGP-2025-09} (right). The two measurements use the same
production-side STXS binning. The markers show the observed values
and the error bars the total uncertainty in the published input cross-sections. The horizontal bars show the
Standard Model predictions of the same references, rescaled by the same factor, and the shaded bands the
uncertainty in those predictions; where this uncertainty is small, the band is hidden by the bar. The Run-2 cross-sections are rescaled with the $\AsfSM$ values obtained from the Run-3 simulation. The qq2Hqq-BSM entry of the Run-3 panel is a 95\% CL upper limit, drawn as a
horizontal bar with a dashed downward arrow. The lower panel shows the ratio of the observed values to the Standard Model
predictions, with the uncertainty in the prediction shown separately as the grey band around unity.}
\label{fig:slif_stage12}
\end{figure}

\Cref{eq:rescale} is applied to every published category with a central value or an upper limit,
using the observed $\sigma_{|\yH|<2.5}\cdot B(H\to ZZ^{*})$ of Ref.~\cite{HIGP-2025-09} and the
per-category SM acceptance $\AsfSM$ of \cref{sec:acceptance}. \Cref{tab:ssf} summarises the results. The rescaled Run-3 cross-sections are compared with the Standard Model predictions in Figures~\ref{fig:slif_stage0} and~\ref{fig:slif_stage12}. The quoted uncertainties are those of the published input cross-sections. The relative binomial uncertainties in $\AsfSM$, ranging from $0.01\%$ to $0.12\%$, are negligible compared with the overall measurement uncertainties of $15$--$100\%$ and are therefore not propagated into the final result. In the case of an asymmetric confidence interval of a published cross-section measurement, both bounds are scaled by the same factor as the central value, such that the rescaled interval retains the original asymmetry. Uncertainties due to acceptance modelling, branching ratios and bin-to-bin covariances are not
included. In the qq2Hqq-BSM STXS production bin, only a $95\%$~CL upper limit has been published. Therefore, no rescaled
central value is quoted.

The same algebraic rescaling is applied to the Run-2 measurement~\cite{HIGG-2018-28}
($139\fbinv$, $\sqrt{s}=13$~\TeV), so that the Run-2 results are also available in the LIF region. Its Production-Mode and STXS
categories map onto the same labels, and the Run-3 acceptances are reused. This is an assumption, but the two are expected to agree. $\AsfSM$ is a particle-level decay quantity built from
invariant four-lepton observables, so it is not expected to depend on the collision energy. The
out-of-fiducial residual should likewise be similar, because the baseline reconstruction selection
is essentially unchanged between the two periods. A dedicated Run-2 evaluation has not been performed here. In both cases the observed, rather than the expected, values of $S_i$ are used. The Run-2 and Run-3 columns of
\cref{tab:ssf} are not expected to coincide, since the cross-sections themselves differ between the
two centre-of-mass energies.

\begin{table}[htbp]
\centering
\footnotesize
\caption{Rescaled LIF cross-sections $\Ssf_i=S_i\cdot B(ZZ^{*}\!\to4\ell)\cdot\AsfSMbin{i}$
for the Run-3 categories of Ref.~\cite{HIGP-2025-09} and, applying the same procedure, for the corresponding categories of the Run-2 measurement of Ref.~\cite{HIGG-2018-28}. $S_i$ is
the published $\sigma_{|\yH|<2.5}\cdot B(H\to ZZ^{*})$ in fb; the same Run-3-based $\AsfSM$ is used for both runs.
Asymmetric published intervals are scaled by the same factor as the central value. The two measurements use the same
production-side STXS binning.}
\label{tab:ssf}
\begin{tabular}{lcr@{}lr@{}lr@{}lr@{}l}
\toprule
& & \multicolumn{4}{c}{Run 3, $165\fbinv$} & \multicolumn{4}{c}{Run 2, $139\fbinv$} \\
\cmidrule(lr){3-6}\cmidrule(lr){7-10}
Category & \AsfSM & \multicolumn{2}{c}{$S_i$ [fb]} & \multicolumn{2}{c}{$\Ssf_i$ [fb]} & \multicolumn{2}{c}{$S_i$ [fb]} & \multicolumn{2}{c}{$\Ssf_i$ [fb]} \\
\midrule
\multicolumn{10}{l}{\textit{Stage-0 / Production-Mode Stage}} \\
ggF                  & $0.8546$ & $1110$ & ${}\pm160$ & $4.5$ & ${}\pm0.6$ & $1120$ & ${}\pm130$ & $4.5$ & ${}\pm0.5$ \\
VBF                  & $0.8545$ & $140$ & ${}\pm50$ & $0.57$ & ${}\pm0.20$ & $110$ & ${}\pm42$ & $0.44$ & ${}\pm0.17$ \\
VH                   & $0.8543$ & $130$ & $^{+69}_{-52}$ & $0.53$ & $^{+0.28}_{-0.21}$ & $75$ & $^{+61}_{-48}$ & $0.30$ & $^{+0.25}_{-0.19}$ \\

Inclusive            & $0.8546$ & $1420$ & ${}\pm160$ & $5.7$ & ${}\pm0.6$ & $1340$ & ${}\pm120$ & $5.4$ & ${}\pm0.5$ \\
\midrule
\multicolumn{10}{l}{\textit{STXS Reduced Stage-1.2}} \\
gg2H-0j-\ptH-Low     & $0.8544$ & $180$ & ${}\pm50$ & $0.73$ & ${}\pm0.20$ & $170$ & ${}\pm54$ & $0.69$ & ${}\pm0.22$ \\
gg2H-0j-\ptH-High    & $0.8547$ & $510$ & ${}\pm100$ & $2.1$ & ${}\pm0.4$ & $630$ & ${}\pm108$ & $2.5$ & ${}\pm0.4$ \\
gg2H-1j-\ptH-Low     & $0.8550$ & $180$ & ${}\pm70$ & $0.73$ & ${}\pm0.28$ & $50$ & $^{+81}_{-92}$ & $0.20$ & $^{+0.33}_{-0.37}$ \\
gg2H-1j-\ptH-Med     & $0.8545$ & $40$ & ${}\pm40$ & $0.16$ & ${}\pm0.16$ & $170$ & $^{+54}_{-51}$ & $0.69$ & $^{+0.22}_{-0.21}$ \\
gg2H-1j-\ptH-High    & $0.8547$ & $13$ & $^{+18}_{-14}$ & $0.05$ & $^{+0.07}_{-0.06}$ & $9$ & $^{+16}_{-11}$ & $0.04$ & $^{+0.06}_{-0.04}$ \\
gg2H-2j              & $0.8543$ & $150$ & ${}\pm70$ & $0.61$ & ${}\pm0.28$ & $40$ & ${}\pm81$ & $0.16$ & ${}\pm0.33$ \\
gg2H-\ptH-High       & $0.8534$ & $12$ & $^{+12}_{-9}$ & $0.05$ & $^{+0.05}_{-0.04}$ & $38$ & $^{+21}_{-16}$ & $0.15$ & $^{+0.08}_{-0.06}$ \\
qq2Hqq-VH            & $0.8541$ & $60$ & $^{+43}_{-33}$ & $0.24$ & $^{+0.17}_{-0.13}$ & $21$ & $^{+38}_{-30}$ & $0.08$ & $^{+0.15}_{-0.12}$ \\
qq2Hqq-VBF           & $0.8545$ & $220$ & $^{+89}_{-72}$ & $0.89$ & $^{+0.36}_{-0.29}$ & $150$ & $^{+54}_{-51}$ & $0.61$ & $^{+0.22}_{-0.21}$ \\
qq2Hqq-BSM           & $0.8546$ & \multicolumn{2}{c}{$<11$ (95\% CL)} & \multicolumn{2}{c}{$<0.044$} & $0.5$ & $^{+11}_{-9}$ & $0.00$ & ${}\pm0.04$ \\
VH-Lep               & $0.8540$ & $25$ & $^{+26}_{-18}$ & $0.10$ & $^{+0.11}_{-0.07}$ & $22$ & $^{+28}_{-18}$ & $0.09$ & $^{+0.11}_{-0.07}$ \\

\bottomrule
\end{tabular}
\end{table}

\FloatBarrier
\section{Conclusion}

The published $H\to ZZ^{*}\to4\ell$ STXS cross-sections have been recast into a Lorentz-invariant fiducial decay region built from invariant four-lepton observables, with no requirement on individual lepton momenta or directions. The acceptance in this fiducial region, based on the Run-3 Higgs boson signal MC simulation, is $\AsfSM=0.8546$. Its spread is $0.07\%$ across different Higgs boson signal samples, $0.04\%$ across Stage-0 modes
and $0.19\%$ across the populated Reduced Stage-1.2 bins.

The reconstruction-to-fiducial mapping is stable at the sub-percent level. Its EFT dependence
reaches at most $0.6\%$ for $\rreco$ for a given single Wilson coefficient, while the corresponding $\rfid$ value
without any fiducial requirement varies by up to $93\%$ over the same range. The SM value $\AsfSM$
is therefore adopted as an EFT-independent rescaling factor, and the residual coefficient dependence of $\rfid$ is
retained as a stated modelling limitation. The fiducial acceptance itself, $\Asfc$, changes by up to $46\%$ over the reference
intervals; this dependence is part of the EFT prediction for the LIF region.

The resulting recast cross-sections can be used as direct inputs to EFT interpretations, in which the EFT dependence of the fiducial acceptance is part of the particle-level signal prediction and no dedicated EFT-dependent corrections for the reconstruction efficiency are needed. The residual EFT dependence of the reconstruction efficiency in the LIF region is small compared with that of the published extrapolation to the inclusive phase space.

\printbibliography

\end{document}